\documentstyle[12pt,epsf]{article}

\begin{document}

\renewcommand{\theequation}{\thesection.\arabic{equation}}
\newcommand{\reseteqnum}{\setcounter{equation}{0}}

\title{
\hfill
\parbox{4cm}{\normalsize
hep-th/0206012}\\
\vspace{2cm}
D-geometric structure of orbifolds
\vspace{2cm}}
\author{Tomomi Muto\thanks{e-mail address:\tt 
tmuto@ms.u-tokyo.ac.jp}
\vspace{1.0cm}\\
{\normalsize\em Graduate School of Mathematical Science,
University of Tokyo}\\
{\normalsize\em Komaba 3-8-1, Meguro-ku, Tokyo 153-8914, Japan}}
\date{\normalsize}
\maketitle

\vspace{1.0cm}

\begin{abstract}
\normalsize
We study D-branes on abelian orbifolds ${\bf C}^d/{\bf Z}_N$ for $d=2, 3$.
The toric data describing the D-brane vacuum moduli space,
which represents the geometry probed by D-branes,
has certain redundancy compared with the classical geometric description of the orbifolds.
We show that the redundancy has a simple combinatorial structure
and find analytic expressions for degrees of the redundancy.
For $d=2$ the structure of the redundancy has a connection with representations of $SU(N)$ Lie algebra,
which provides a new correspondence between geometry and representation theory.
We also prove that non-geometric phases do not appear in the K$\ddot{\rm a}$hler moduli space for $d=2$.
\end{abstract}

\newpage
\section{Introduction}

D-geometry, the geometry as seen by D-branes,
has qualitatively different features from that probed by fundamental strings.
One of the remarkable features of the D-geometry is that
non-geometric phases, in which the orbifold singularity is resolved,
are projected out from the K$\ddot{\rm a}$hler moduli space \cite{DGM, Muto}
in contrast to the analysis based on fundamental strings \cite{AGM, Witten}.
By inspecting the calculation in \cite{DGM},
one can see that the projection of the non-geometric phases stems from certain redundancy
of lattice vectors describing the toric data of the orbifold:
the redundancy of the  lattice vectors implies  redundancy of coordinates describing the toric variety,
and elimination of the redundant variables leads to the resolution of the orbifold singularity.
Therefore it is important to clarify the structure of the redundancy to understand
the nature of D-geometry.
The redundancy, however, occurs as a result of a combinatorial algorithm
based on toric geometry, and even the degree of the redundancy has not been known except for 
the models for which explicit calculation was carried out\footnote{The meaning of the redundancy
for the orbifold ${\bf C}^3/{\bf Z}_2 \times {\bf Z}_2$ is discussed in \cite{MT}.}.
For example, the orbifolds of the form
${\bf C}^3/{\bf Z}_N$, ($(x,y,z) \equiv  (\omega x,\omega y,\omega^{-2}z)$ with $\omega^N=1$, $N$:odd)
have been analyzed only for $N \leq 11$ \cite{Muto}.
Total number $n_D(N)$ of lattice vectors in the toric data which describes the D-geometry of the orbifold
are shown in table \ref{table:number}.
For reference we also show the total number $n_C(N)$ of lattice vectors
in the classical geometric description of the orbifold.
As one easily see, $n_C(2N+1)=N+3$,
while the analytic expression for the number $n_D(N)$ has not been known. 
\begin{table}[h]
\begin{center}
\label{table:number}
\begin{tabular}{|c|c|c|c|c|c|}
\hline
$N$&3&5&7&9&11 \\ \hline
$n_C(N)$&4&5&6&7&8 \\ \hline
$n_D(N)$&6&13&31&78&201 \\ \hline
\end{tabular}
\caption{$n_D(N)$ is the total number of lattice vectors
necessary to describe the D-geometry the orbifold ${\bf C}^3/{\bf Z}_N$.
$n_C(N)$ is the total number of lattice vectors
necessary to describe the orbifold ${\bf C}^3/{\bf Z}_N$ in classical geometry.}
\end{center}
\end{table}

The purpose of the present paper is to clarify the structure of the redundancy
by re-examining the toric data carefully.
In section 2 we investigate two-dimensional abalian orbifold ${\bf C}^2/{\bf Z}_3$
with ${\bf Z}_3 \in SU(2)$ to illustrate the idea of our analysis.
We find that the structure of the redundancy is related to representations of $SU(3)$
Lie algebra.
It implies that there is a correspondence between exceptional divisors
of the resolution of the orbifold ${\bf C}^2/{\bf Z}_3$
and representations of the $SU(3)$ Lie algebra.
We discuss the meaning of the elimination of the redundant variables from the viewpoint of representation theory.
In section 3, we generalize the analysis to the orbifold ${\bf C}^2/{\bf Z}_N$
and find a relation between the D-geometry of ${\bf C}^2/{\bf Z}_N$ and $SU(N)$ Lie algebra.
We also show that non-geometric phases are projected out for any $N$.
In section 4, we study the three-dimensional abelian orbifold mentioned above.
We show that there is a simple combinatrial structure similar to the two-dimensional case
and derive the analytic expression for the number $n_D(N)$.

After completion of this work, a paper \cite{FFHH} appeared which has overlap with ours.

\section{D-geometric structure of ${\bf C}^2/{\bf Z}_3$}
\reseteqnum

\begin{figure}[htp]
\begin{center}
\begin{picture}(200,100)
\put(60,60){$x_1$}
\put(76,44){$y_1$}
\put(97,8){$x_2$}
\put(97,28){$y_2$}
\put(129,60){$x_3$}
\put(113,44){$y_3$}
\put(40,20){\circle{10}}
\put(160,20){\circle{10}}
\put(100,80){\circle{10}}
\put(50,17){\vector(1,0){100}}
\put(150,23){\vector(-1,0){100}}
\put(50,25){\vector(1,1){45}}
\put(90,75){\vector(-1,-1){45}}
\put(105,70){\vector(1,-1){45}}
\put(155,30){\vector(-1,1){45}}
\end{picture}
\caption{The quiver diagram for ${\bf C}^2/{\bf Z}_3$.}
\label{fig:quiverC2Z3}
\end{center}
\end{figure}
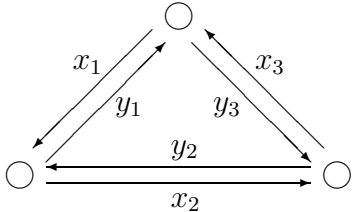

In this section we study D-branes on the orbifold ${\bf C}^2/{\bf Z}_3$
where the action of $g \in {\bf Z}_3$ on $(x,y) \in {\bf C}^2$ is given by
\begin{equation}
(x, y) \rightarrow (\omega_3 x, \omega_3^{-1}y), \quad \omega_3={\rm exp}(2\pi i/3).
\end{equation}
Field contents of the worldvolume gauge theory of the D-brane
is encoded in the quiver diagram \cite{DM}.
The quiver diagram of this model is depicted in Figure \ref{fig:quiverC2Z3}.
Each node represents a factor $U(1)$ of the gauge group and each arrow represents a complex
scalar field.
Note that the gauge group is $U(1)^2$ since the diagonal $U(1)$ of $U(1)^3$ is trivial. 

The classical moduli space $\cal M$ of the quiver gauge theory,
which is interpreted as the space probed by the D-brane,
is obtained from the space ${\bf C}^6$ of complex scalars $(x_i,y_i)$ 
by imposing the F-flatness conditions
\begin{equation}
x_1 y_1=x_2 y_2=x_3 y_3,
\label{eq:F-flat3}
\end{equation}
and D-flatness conditions
\begin{eqnarray}
&&|x_1|^2-|y_1|^2-|x_2|^2+|y_2|^2 = \zeta_1,\nonumber \\
&&|x_2|^2-|y_2|^2-|x_3|^2+|y_3|^2 = \zeta_2,
\label{eq:D-flat3}
\end{eqnarray}
and further dividing by the gauge group $U(1)^2$.
The Fayet-Iliopoulos parameters $\zeta_i$ coming from twisted sector of closed strings
parameterize the K$\ddot{\rm a}$hler moduli space of the orbiofld.

We calculate the space $\cal M$
following the procedure given in \cite{DGM}.
We first consider the space $\cal W$ on which only the F-flatness constraints (\ref{eq:F-flat3})
are imposed,
\begin{equation}
{\cal W}=\{(x_i,y_i) \in {\bf C}^6 | x_1 y_1=x_2 y_2=x_3 y_3 \}.
\end{equation}
As explained in \cite{DGM},
$\cal W$ can be expressed as a holomophic quotient of the form
\begin{equation}
({\bf C}^k-F)/({\bf C}^*)^{k-4},
\label{eq:holomorphic}
\end{equation}
and furthermore it is realized as a vacuum moduli space of
a certain two-dimensional $N=(2,2)$ supersymmetric $U(1)^{k-4}$ gauged linear sigma model \cite{Witten}.
Homogeneous coordinates $p_i$ of ${\bf C}^k$ is interpreted as scalar components
of chiral superfields $P_i$ in the gauge theory,
and the ${\bf C}^*$ quotient is realized by the combined operation of imposing D-flatness constraint
and then dividing by the $U(1)$ gauge symmetry.
The procedure for realizing $\cal W$ as a vacuum moduli space of a gauge theory
is carried out by a combinatorial algorithm based on toric geometry.
The first step of the algorithm is to write the solution of the F-flatness constraints as
\begin{eqnarray}
&&x_1=u_1, \quad y_1=u_2, \quad y_2=u_3, \quad y_3=u_4, \nonumber \\
&&x_2=u_1 u_2 u_3^{-1}, \quad x_3=u_1 u_2 u_4^{-1}
\end{eqnarray}
by introducing four complex variables $u_1, \cdots, u_4$.
(Note that the dimensions of $\cal W$ is four.)
The form of the solution
is specified by the matrix
\begin{equation}
K=\left(
\begin{array}{lcccc}
&u_1&u_2&u_3&u_4\\
m_1 \, &1&0&0&0\\
m_2 \, &0&1&0&0\\
m_3 \, &0&0&1&0\\
m_4 \, &0&0&0&1\\
m_5 \, &1&1&-1&0\\
m_6 \, &1&1&0&-1\\
\end{array}
\right).
\end{equation}
The row vectors $m_i$ of the matrix $K$
define the edges of a cone $\hat \sigma$ in ${\bf R}^4$.
The next step is to calculate the dual cone of $\hat \sigma$ defined by
\begin{equation}
\sigma =\{n \in {\bf R}^4 | m \cdot n \geq 0, \forall m \in \hat \sigma\}.
\end{equation}
Calculation of $\sigma$ is a problem of integer programming,
and in the present case
the cone $\sigma$ is generated by eight vectors $n_i$ in the lattice ${\bf Z}^4$.
They are represented by column vectors in the matrix,
\begin{equation}
T=\left(
\begin{array}{lcccccccc}
&n_1&n_2&n_3&n_4&n_5&n_6&n_7&n_8\\
u_1&1&0&1&1&0&0&1&0\\
u_2&0&1&0&0&1&1&0&1\\
u_3&0&0&1&0&1&0&1&1\\
u_4&0&0&0&1&0&1&1&1
\label{eq:cone}
\end{array}
\right).
\end{equation}
Chiral multiplets $P_i$ in the gauge theory realization
corresponds to each vector $n_i$,
so the number $k$ in (\ref{eq:holomorphic}) is equal to 8.
D-flatness conditions imposed on the variables $p_i$ are
determined from linear relations among the vectors $n_i$.
The four linear relations
\begin{eqnarray}
&&n_1-n_2-n_3+n_5=0, \nonumber \\
&&n_1-n_3-n_4+n_7=0, \nonumber \\
&&n_1-n_4-n_2+n_6=0, \\
&&n_4-n_6-n_7+n_8=0, \nonumber
\end{eqnarray}
imply four D-flatness conditions
in the gauge theory realization,
\begin{eqnarray}
&&|p_1|^2-|p_2|^2-|p_3|^2+|p_5|^2=0, \nonumber \\
&&|p_1|^2-|p_3|^2-|p_4|^2+|p_7|^2=0, \nonumber \\
&&|p_1|^2-|p_4|^2-|p_2|^2+|p_6|^2=0, \label{eq:FD} \\
&&|p_4|^2-|p_6|^2-|p_7|^2+|p_8|^2=0. \nonumber
\end{eqnarray}
Thus $\cal W$ is obtained from the space ${\bf C}^8$ with coordinates $(p_1,\cdots,p_8)$
by imposing the four D-flatness conditions (\ref{eq:FD}) and dividing by the $U(1)^4$ gauge symmetry.

To combine the D-flatness conditions ($\ref{eq:D-flat3}$)
existing from the beginning,
we must represent the equations ($\ref{eq:D-flat3}$) in terms of the variables $p_i$.
Since the relation between the coordinates
$(v_1,\cdots,v_6)=(x_1,y_1,y_2,y_3,x_2,x_3)$ and
$(p_1,\cdots,p_8)$ is given by
\begin{equation}
v_i=\prod_j p_j^{m_i \cdot n_j},
\end{equation}
the D-flatness conditions are rewritten as
\begin{eqnarray}
&&|p_2|^2-|p_3|^2=\zeta_1,\nonumber \\
&&|p_3|^2-|p_4|^2=\zeta_2. 
\label{eq:DD}
\end{eqnarray}
Thus the moduli space $\cal M$ is described by the six D-flatness conditions
(\ref{eq:FD}) and (\ref{eq:DD}).
For later use we make some rearrangements,
\begin{eqnarray}
&&|p_1|^2-|p_3|^2-|p_4|^2+|p_7|^2=0, \nonumber \\
&&|p_4|^2-|p_6|^2-|p_7|^2+|p_8|^2=0, \nonumber \\
&&|p_2|^2-|p_3|^2=\zeta_1, \nonumber \\
&&|p_3|^2-|p_4|^2=\zeta_2, \label{eq:D} \\
&&|p_5|^2-|p_6|^2=\zeta_2, \nonumber \\
&&|p_6|^2-|p_7|^2=\zeta_1. \nonumber
\end{eqnarray}

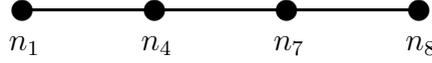
\begin{figure}[htp]
\begin{center}
\begin{picture}(250,30)
\put(45,5){$n_1$}
\put(95,5){$n_4$}
\put(145,5){$n_7$}
\put(195,5){$n_8$}
\put(50,20){\circle*{8}}
\put(100,20){\circle*{8}}
\put(150,20){\circle*{8}}
\put(200,20){\circle*{8}}
\thicklines
\put(50,20){\line(1,0){150}}
\end{picture}
\caption{Toric diagram representing $\cal M$ in the region $\zeta_1>0$ and $\zeta_2>0$.}
\label{fig:toric3}
\end{center}
\end{figure}

The analysis of $\cal M$ described by (\ref{eq:D}) depends on the values of $(\zeta_1,\zeta_2)$.
We first examine in the region $\zeta_1>0$ and $\zeta_2>0$,
where we obtain inequalities $|p_2|^2>|p_3|^2>|p_4|^2$ and $|p_5|^2>|p_6|^2>|p_7|^2$
from the last four equations of (\ref{eq:D}).
Hence the four variables $p_2$, $p_3$, $p_5$ and $p_6$ are not zero in this region.
In such a situation, the four variables can be eliminated by using the last four D-flatness conditions of (\ref{eq:D})
and the corresponding $U(1)^4$ gauge symmetry.
Thus we obtain two D-flatness conditions on four variables $p_1$, $p_4$, $p_7$ and $p_8$,
\begin{eqnarray}
&&|p_1|^2-2|p_4|^2+|p_7|^2=\zeta_2, \nonumber \\
&&|p_4|^2-2|p_7|^2+|p_8|^2=\zeta_1.
\end{eqnarray}
Since we are considering in the region $\zeta_1>0$ and $\zeta_2>0$,
the right hand side of the equations are positive,
which implies that $\cal M$ is the resolution of the orbifold ${\bf C}^2/{\bf Z}_3$.
The topology of $\cal M$ is represented by the toric diagram in Figure \ref{fig:toric3}.
In the figure, the vector $n_i$ corresponding to the variable $p_i$ is attached to each vertex.

Next we examine (\ref{eq:D}) in the region $\zeta_1<0$ and $\zeta_1+\zeta_2>0$ as another example.
Then we obtain inequalities $|p_3|^2>|p_2|^2>|p_4|^2$ and $|p_5|^2>|p_7|^2>|p_6|^2$
from the last four equations of (\ref{eq:D}),
and hence the four variables $p_2$, $p_3$, $p_5$ and $p_7$ are not zero.
In such a situation, the four variables are redundant.
Thus we obtain two D-flatness conditions on four variables $p_1$, $p_4$, $p_6$ and $p_8$,
\begin{eqnarray}
&&|p_1|^2-2|p_4|^2+|p_6|^2=\zeta_1+\zeta_2, \nonumber \\
&&|p_4|^2-2|p_6|^2+|p_8|^2=-\zeta_1.
\end{eqnarray}
Although the variable $p_7$ is replaced by $p_6$ compared with the first case,
the form of the equations are the same
and furthermore the right hand side of the two D-flatness equations remain to be positive
since we are now considering in the region $\zeta_1<0$ and $\zeta_1+\zeta_2>0$.
Hence $\cal M$ is the resolution of the orbifold ${\bf C}^2/{\bf Z}_3$ also in this region.
These two examples show that the exchange of variables from $p_7$ to $p_6$
is the key to the fact that
$\cal M$ remains to be the resolution of the orbifold ${\bf C}^2/{\bf Z}_3$:
if such an exchange of variables does not occur,
$\cal M$ becomes singular after crossing the line $\zeta_1=0$.

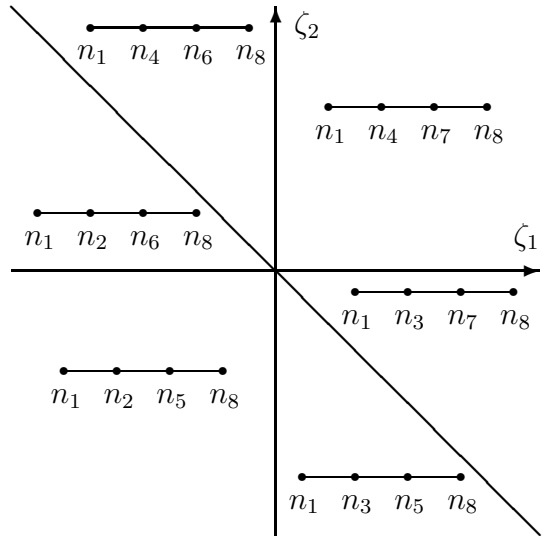
\begin{figure}[htp]
\begin{center}
\begin{picture}(200,200)
\put(120,162){\line(1,0){60}}
\put(120,162){\circle*{3}}
\put(140,162){\circle*{3}}
\put(160,162){\circle*{3}}
\put(180,162){\circle*{3}}
\put(115,150){$n_1$}
\put(135,150){$n_4$}
\put(155,150){$n_7$}
\put(175,150){$n_8$}
\put(30,192){\line(1,0){60}}
\put(30,192){\circle*{3}}
\put(50,192){\circle*{3}}
\put(70,192){\circle*{3}}
\put(90,192){\circle*{3}}
\put(25,180){$n_1$}
\put(45,180){$n_4$}
\put(65,180){$n_6$}
\put(85,180){$n_8$}
\put(10,122){\line(1,0){60}}
\put(10,122){\circle*{3}}
\put(30,122){\circle*{3}}
\put(50,122){\circle*{3}}
\put(70,122){\circle*{3}}
\put(05,110){$n_1$}
\put(25,110){$n_2$}
\put(45,110){$n_6$}
\put(65,110){$n_8$}
\put(20,62){\line(1,0){60}}
\put(20,62){\circle*{3}}
\put(40,62){\circle*{3}}
\put(60,62){\circle*{3}}
\put(80,62){\circle*{3}}
\put(15,50){$n_1$}
\put(35,50){$n_2$}
\put(55,50){$n_5$}
\put(75,50){$n_8$}
\put(110,22){\line(1,0){60}}
\put(110,22){\circle*{3}}
\put(130,22){\circle*{3}}
\put(150,22){\circle*{3}}
\put(170,22){\circle*{3}}
\put(105,10){$n_1$}
\put(125,10){$n_3$}
\put(145,10){$n_5$}
\put(165,10){$n_8$}
\put(130,92){\line(1,0){60}}
\put(130,92){\circle*{3}}
\put(150,92){\circle*{3}}
\put(170,92){\circle*{3}}
\put(190,92){\circle*{3}}
\put(125,80){$n_1$}
\put(145,80){$n_3$}
\put(165,80){$n_7$}
\put(185,80){$n_8$}
\thicklines
\put(0,100){\vector(1,0){200}}
\put(100,0){\vector(0,1){200}}
\put(0,200){\line(1,-1){200}}
\put(190,110){$\zeta_1$}
\put(107,190){$\zeta_2$}
\end{picture}
\caption{Phase structure for ${\bf C}^2/{\bf Z}_3$.}
\label{fig:phase}
\end{center}
\end{figure}

Similar analysis shows that the equations (\ref{eq:D}) describe
the resolution of the orbifold ${\bf C}^2/{\bf Z}_3$
independent of the region in the $(\zeta_1,\zeta_2)$ space
by choosing appropriate set of variables,
thus non-geometric phases are projected out from the the K$\ddot{\rm a}$hler moduli space 
of the orbifold\footnote{On
codimension one loci in the $\zeta$-space,
the singularity does not completely resolved at the classical level.}.
In Figure \ref{fig:phase}
we summarize toric diagrams representing $\cal W$ and lattice vectors $n_i$ corresponding to the
variables $p_i$ chosen in each region in the $\zeta$-space.

As shown in the Figure, the vector $n_1$ ($n_8$) corresponds to the left (right) vertex
independent of the values of $(\zeta_1,\zeta_2)$.
On the other hand, the three vectors $n_2,n_3,n_4$ ($n_5,n_6,n_7$)
correspond to the second vertex from the left (right).
One of the three vectors is chosen in each region in the $\zeta$-space and
the other two are redundant.

\begin{figure}[htp]
\begin{center}
\begin{picture}(400,100)
\put(20,35){$\hat n_1=(0,0,0)$}
\put(120,50){$\hat n_2=(1,0,0)$}
\put(120,35){$\hat n_3=(0,1,0)$}
\put(120,20){$\hat n_4=(0,0,1)$}
\put(220,50){$\hat n_5=(1,1,0)$}
\put(220,35){$\hat n_6=(1,0,1)$}
\put(220,20){$\hat n_7=(0,1,1)$}
\put(320,35){$\hat n_8=(1,1,1)$}
\put(50,80){\circle*{8}}
\put(150,80){\circle*{8}}
\put(250,80){\circle*{8}}
\put(350,80){\circle*{8}}
\thicklines
\put(50,80){\line(1,0){300}}
\end{picture}
\caption{Toric diagram for ${\bf C}^2/{\bf Z}_3$.}
\label{fig:toricC2Z3}
\end{center}
\end{figure}
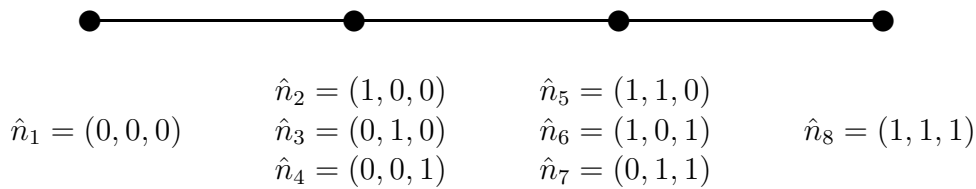

To understand the meaning of the redundancy,
let us look closely at the vectors $n_i$ in ($\ref{eq:cone}$).
We first note that every vector $n_i$ satisfies the following condition
\begin{equation}
n_i \cdot \rho_0 =1, \quad {\rm for} \quad \rho_0=(1,1,0,0),
\end{equation}
which corresponds to the fact that the space ${\cal M}$ satisfies Calabi-Yau condition.
Thus $n_i$ lies on a three-dimensional hyperplane,
and it is enough to consider three-dimensional vectors $\hat n_i$ obtained by omitting
the first entry of $n_i$.
The vectors $\hat n_i$ corresponding to each vertex of the toric diagram
are given in Figure \ref{fig:toricC2Z3}.

Structure of the redundancy is clear from the form of the vectors;
the three entries of each vector are equal to 0 or 1,
and vectors whose $k$ components are equal to 1
correspond to the $k+1$-th vertex from the left.
Hence the multiplicity of the vectors associated to the $k+1$-th vertex from the left
is given by $(^3_k)$.

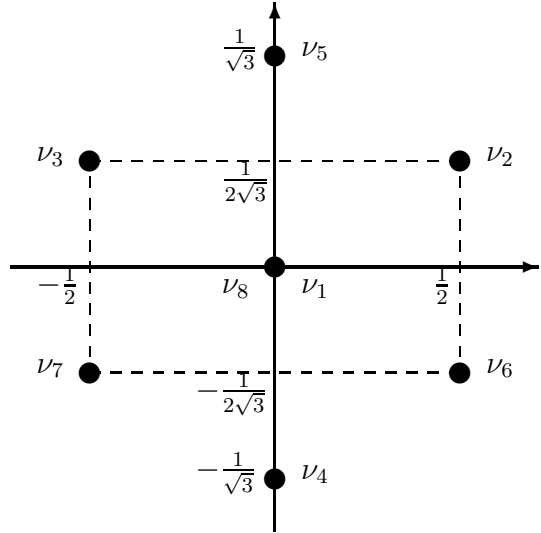
\begin{figure}[htp]
\begin{center}
\begin{picture}(200,200)
\put(30,60){\dashbox{4}(140,80)}
\put(100,100){\circle*{8}}
\put(170,140){\circle*{8}}
\put(30,140){\circle*{8}}
\put(100,20){\circle*{8}}
\put(100,180){\circle*{8}}
\put(170,60){\circle*{8}}
\put(30,60){\circle*{8}}
\put(110,90){$\nu_1$}
\put(180,140){$\nu_2$}
\put(10,140){$\nu_3$}
\put(110,20){$\nu_4$}
\put(110,180){$\nu_5$}
\put(180,60){$\nu_6$}
\put(10,60){$\nu_7$}
\put(80,90){$\nu_8$}
\put(160,90){$\frac{1}{2}$}
\put(10,90){$-\frac{1}{2}$}
\put(70,20){$-\frac{1}{\sqrt 3}$}
\put(80,180){$\frac{1}{\sqrt 3}$}
\put(70,50){$-\frac{1}{2\sqrt 3}$}
\put(80,130){$\frac{1}{2\sqrt 3}$}
\thicklines
\put(0,100){\vector(1,0){200}}
\put(100,0){\vector(0,1){200}}
\end{picture}
\caption{Eigenvalues of the vectors $\hat n_i$.}
\label{fig:weight}
\end{center}
\end{figure}

\begin{figure}[htp]
\begin{center}
\begin{picture}(400,60)
\put(47,27){1}
\put(145,25){\framebox(10,10)}
\put(245,15){\framebox(10,10)}
\put(245,25){\framebox(10,10)}
\put(345,5){\framebox(10,10)}
\put(345,15){\framebox(10,10)}
\put(345,25){\framebox(10,10)}
\put(50,50){\circle*{8}}
\put(150,50){\circle*{8}}
\put(250,50){\circle*{8}}
\put(350,50){\circle*{8}}
\thicklines
\put(50,50){\line(1,0){300}}
\end{picture}
\caption{Correspondence between representations of $SU(3)$
and the toric diagram for ${\bf C}^2/{\bf Z}_3$.
The Dynkin diagrams indicate the representations.}
\label{fig:representationC2Z3}
\end{center}
\end{figure}
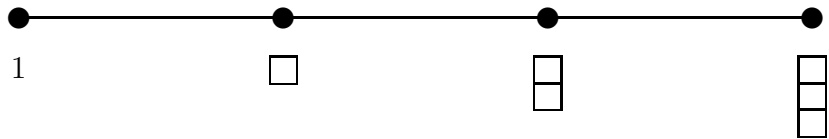

Now we would like to show that the above structure has correspondence
with representations of the $SU(3)$ Lie algebra.
Let us consider the eigenvalues $\nu_i$ of the vectors $\hat n_i$
for the generators of the Cartan subalgebra of $SU(3)$,
\begin{equation}
\frac{1}{2}
\left(
\begin{array}{ccc}
1&0&0 \\
0&-1&0 \\
0&0&0
\end{array}
\right),\quad
\frac{1}{2\sqrt 3}
\left(
\begin{array}{ccc}
1&0&0 \\
0&1&0 \\
0&0&-2
\end{array}
\right).
\end{equation}
We obtain the following results,
\begin{eqnarray}
&&\nu_1=(0,0),\\
&&\nu_2=\left(\frac{1}{2},\frac{1}{2\sqrt 3}\right),
\nu_3=\left(-\frac{1}{2},\frac{1}{2\sqrt 3}\right),
\nu_4=\left(0,-\frac{1}{\sqrt 3}\right),\\
&&\nu_5=\left(0,\frac{1}{\sqrt 3}\right),
\nu_6=\left(\frac{1}{2},-\frac{1}{2\sqrt 3}\right),\nu_7=\left(-\frac{1}{2},-\frac{1}{2\sqrt 3}\right),\\
&&\nu_8=(0,0),
\end{eqnarray}
and they are depicted in Figure \ref{fig:weight}.
The figure implies that the three vectors $\nu_2$, $\nu_3$ and $\nu_4$ form the weights of the representation
${\bf 3}$ of $SU(3)$
and the three vectors $\nu_5$, $\nu_6$ and $\nu_7$ form the weights of the representation
$\bar {\bf 3}$ of SU(3).
Thus the three vectors $n_2,n_3,n_4$ ($n_5,n_6,n_7$)
associated to the second vertex from the left (right) in the toric diagram
correspond to the representation ${\bf 3}$ ($\bar {\bf 3}$) of $SU(3)$.
Furthermore it is reasonable to consider the vector $n_1$ ($n_8$)
associated to the left (right) vertex in the toric diagram
corresponds to the representation
${\bf 1}$ ($\bar {\bf 1}$) of $SU(3)$,
where ${\bf 1}$ is the trivial representation.
Thus antisymmetric representations of $SU(3)$ corresponds to each vertex in the toric diagram
as depicted in Figure \ref{fig:representationC2Z3}.
The degrees of the redundancy (the multiplicity of the vectors)
coincide with the dimensions of the representations.

Since the vertices except for the both ends of the diagram represent exceptional divisors,
the graph implies that there is a correspondence between exceptional divisors
of the resolution of the orbifold ${\bf C}^2/{\bf Z}_3$
and antisymmetric representations of the Lie algebra $SU(3)$.
Note that the McKay correspondence \cite{McKay}
gives the connection between exceptional divisors of the resolution of ${\bf C}^2/\Gamma$
and irreducible representations of $\Gamma$.
The above correspondence gives a new kind of connection between geometry and representation theory.

Finally we would like to discuss positivity of roots of the $SU(3)$ Lie algebra.
It gives a simple explanation on the choice of four variables $p_i$ in each
region in the $\zeta$-space.
In identifying the degeneracy of the lattice vectors $n_i$ with representations of 
$SU(3)$,
$\nu_i-\nu_j$ is identified with a root of the $SU(3)$ Lie algebra.
However, the positivity of roots is irrelevant in the discussion.
Now we define the positivity of the roots as
\begin{eqnarray}
&&sign(\nu_2-\nu_3)=sign (\zeta_1), \nonumber \\
&&sign(\nu_3-\nu_4)=sign (\zeta_2), \\
&&sign(\nu_2-\nu_4)=sign (\zeta_1+\zeta_2).\nonumber
\end{eqnarray}
The definition comes from the D-flatness conditions
\begin{eqnarray}
&&|p_2|^2-|p_3|^2=\zeta_1, \nonumber \\
&&|p_3|^2-|p_4|^2=\zeta_2.
\end{eqnarray}
The conditions lead to the relations
\begin{eqnarray}
&&sign(\nu_5-\nu_6)=sign (\zeta_2) \nonumber \\
&&sign(\nu_6-\nu_7)=sign (\zeta_1)\\
&&sign(\nu_5-\nu_7)=sign (\zeta_1+\zeta_2),\nonumber
\end{eqnarray}
which corresponds to the D-flatness conditions
\begin{eqnarray}
&&|p_5|^2-|p_6|^2=\zeta_2, \nonumber \\
&&|p_6|^2-|p_7|^2=\zeta_1.
\end{eqnarray}

By definition, positivity depend on the values of $(\zeta_1,\zeta_2)$.
In the region $\zeta_1>0$ and $\zeta_2>0$, for example,
we obtain inequalities
\begin{eqnarray}
&&\nu_2>\nu_3>\nu_4, \nonumber \\
&&\nu_5>\nu_6>\nu_7,
\end{eqnarray}
They mean that $\nu_4$ and $\nu_7$ are lowest weight states
of the representations ${\bf 3}$ and $\bar {\bf 3}$ respectively.
Thus the coordinate $p_i$ which remains after elimination of the redundant variables
corresponds to the lowest weight state in each representation.
This property also holds for any region in the $\zeta$-space.

\section{D-geometric structure of ${\bf C}^2/{\bf Z}_N$}
\reseteqnum

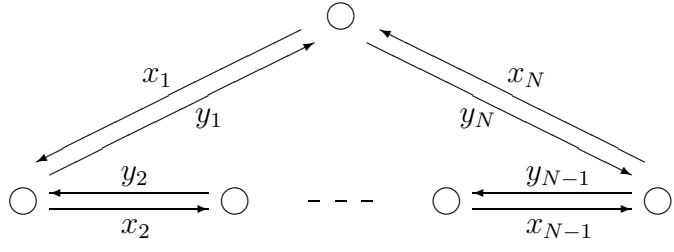
\begin{figure}[htp]
\begin{center}
\begin{picture}(400,100)
\put(125,65){$x_1$}
\put(145,50){$y_1$}
\put(117,8){$x_2$}
\put(117,28){$y_2$}
\put(270,8){$x_{N-1}$}
\put(270,28){$y_{N-1}$}
\put(263,65){$x_N$}
\put(245,50){$y_N$}
\put(80,20){\circle{10}}
\put(160,20){\circle{10}}
\put(240,20){\circle{10}}
\put(320,20){\circle{10}}
\put(200,90){\circle{10}}
\put(90,17){\vector(1,0){60}}
\put(150,23){\vector(-1,0){60}}
\put(250,17){\vector(1,0){60}}
\put(310,23){\vector(-1,0){60}}
\put(90,30){\vector(2,1){100}}
\put(185,85){\vector(-2,-1){100}}
\put(210,80){\vector(2,-1){100}}
\put(315,35){\vector(-2,1){100}}
\put(188,20){\line(1,0){4}}
\put(198,20){\line(1,0){4}}
\put(208,20){\line(1,0){4}}
\end{picture}
\caption{Quiver diagram for ${\bf C}^2/{\bf Z}_N$.}
\label{fig:quiverC2ZN}
\end{center}
\end{figure}

The discussion in the last section is generalized to the orbifold ${\bf C}^2/{\bf Z}_N$
where the action of $g \in {\bf Z}_N$ on $(x,y) \in {\bf C}^2$ is given by
\begin{equation}
(x, y) \rightarrow (\omega_N x, \omega_N^{-1}y), \quad \omega_N={\rm exp}(2\pi i/N).
\end{equation}
The quiver diagram of this model is depicted in Figure \ref{fig:quiverC2ZN}.
The classical moduli space $\cal M$ of the corresponding quiver gauge theory is
obtained from the space ${\bf C}^{2N}$ of variables $(x_i,y_i)$ 
by imposing the F-flatness conditions
\begin{equation}
x_1 y_1=x_2 y_2 = \cdots =x_N y_N,
\label{eq:F-flatN}
\end{equation}
and D-flatness conditions
\begin{equation}
|x_i|^2-|y_i|^2-|x_{i+1}|^2+|y_{i+1}|^2 = \zeta_i, \quad (i=1,2,\cdots,N-1),
\label{eq:D-flatN}
\end{equation}
and further dividing by the gauge group $U(1)^{N-1}$.

We first consider the space 
$\cal W$ imposed only the F-flatness constraints (\ref{eq:F-flatN}).
The solution of the constraints (\ref{eq:F-flatN}) of the form
\begin{eqnarray}
&&x_1=u_1, \quad y_1=u_2, \quad y_2=u_3, \quad \cdots, \quad y_N=u_{N+1}, \nonumber \\
&&x_2=u_1 u_2 u_3^{-1}, \quad x_3=u_1 u_2 u_4^{-1}, \quad \cdots, \quad x_N=u_1 u_2 u_{N+1}^{-1},
\end{eqnarray}
is encoded in the rows of the matrix $K$,
\begin{equation}
K=\left(
\begin{array}{lcccccc}
&u_1&u_2&u_3&u_4&\cdots&u_{N+1}\\
m_1&1&0&0&0&\cdots&0\\
m_2&0&1&0&0&\cdots&0\\
m_3&0&0&1&0&\cdots&0\\
m_4&0&0&0&1&\cdots&0\\
&&&\vdots&&&\\
m_{N+1}&0&0&0&0&\cdots&1\\
m_{N+2}&1&1&-1&0&\cdots&0\\
m_{N+3}&1&1&0&-1&\cdots&0\\
&&&\vdots&&&\\
m_{2N}&1&1&0&0&\cdots&-1
\end{array}
\right).
\end{equation}
The $N+1$-dimensional vectors $m_i$ form a cone $\hat \sigma$ in ${\bf R}^{N+1}$.
To represent the space ${\cal W}$ as a holomorphic quotient,
we consider the dual cone $\sigma$ defined by
\begin{equation}
\sigma =\{n \in {\bf R}^{N+1} | m \cdot n \geq 0, \forall m \in \hat \sigma\}.
\label{eq:dual}
\end{equation}
After some calculations, we obtain the cone $\sigma$ generated by
$2^N$ vectors $n_i$ $(i=1,\cdots,2^N)$ of the form
\begin{equation}
(1,0,*,*,\cdots,*) \quad {\rm or} \quad (0,1,*,*,\cdots,*),
\label{eq:generators}
\end{equation}
where each of the last $N-1$ entries
denoted by $*$ is equal to 0 or 1.
Since the $N+1$-dimensional vector $n_i$ satisfies the relation
$\rho_0 \cdot n_i=1$ for $\rho_0=(1,1,0,\cdots,0)$,
$n_i$ lies on a $N$-dimensional hyperplane,
so we consider $N$-dimensional vectors $\hat n_i$ obtained by omitting
the first entry of $n_i$.
As one can see from (\ref{eq:generators}),
every entry of $\hat n_i$ is equal to 0 or 1,
hence the $2^N$ vectors $\hat n_i$ are labeled by the elements of ${\bf Z}_2^N$;
in other words,
$\hat n_i$ corresponds to a vertex of the $N$-dimensional cube with volume one.

For later use, we define degree $d(\hat n_i)$ of a vector $\hat n_i$ by $d(\hat n_i) =\hat \rho_1 \cdot \hat n_i$
for an $N$-dimensional vector $\hat \rho_1=(1,1,\cdots,1)$.
By definition, the degree $d(\hat n_i)$ is the number of components of $\hat n_i$
whose entry is 1,
and $d(\hat n_i)$ takes values from 0 to $N$.
Thus the vectors ${\hat n_i}$ are classified according to their degrees.
We denote the set of vectors with degree $k$ as $F_k$,
in which there are $(^N_{\, k} )$ elements.
We also introduce $N$-dimensional vectors $e_i$ $(i=1,\cdots,N)$ defined by
\begin{equation}
e_i=(0,\cdots,0,1,0,\cdots,0)
\end{equation}
where the $i$-th entry is equal to 1.
The vectors in $F_k$ are written in the form
\begin{equation}
\hat n_{(a_1,\cdots,a_k)}=e_{a_1}+e_{a_2}+\cdots+e_{a_k}
\end{equation}
where $(a_1,a_2,\cdots,a_N)$ is a permutation of $(1,2,\cdots,N)$.
For $k=0$, we define $\hat n_{(\phi)}=(0,\cdots,0)$.
Note that the vectors $n_i$ and the corresponding complex scalars $p_i$
are also labeled by the same indices.

Under these definitions,
$2^N-N-1$ linear relations among the $2^N$ vectors $n_i$ are
written in the form
\begin{equation}
n_{(a_1,\cdots,a_k)}-n_{(a_1,\cdots,a_k,a_{k+1})}
-n_{(a_1,\cdots,a_k,a_{k+2})}+n_{(a_1,\cdots,a_k,a_{k+1},a_{k+2})}=0,
\label{eq:relation}
\end{equation}
where $k=0,1,\cdots,N-2$.
These relations imply $2^N-N-1$ D-flatness conditions,
\begin{equation}
|p_{(a_1,\cdots,a_k)}|^2-|p_{(a_1,\cdots,a_k,a_{k+1})}|^2
-|p_{(a_1,\cdots,a_k,a_{k+2})}|^2+|p_{(a_1,\cdots,a_k,a_{k+1},a_{k+2})}|^2=0.
\label{eq:FDN}
\end{equation}
These are the D-flatness conditions converted from the F-flatness conditions (\ref{eq:F-flatN}).
On the other hand,
the D-flatness conditions (\ref{eq:D-flatN}) existing from the beginning are rewritten as
\begin{equation}
|p_{(i)}|^2-|p_{(i+1)}|^2=\zeta_i
\label{eq:DDN}
\end{equation}
where $i=1,2,\cdots,N-1$.
Thus the vacuum moduli space $\cal M$ is described by the $2^N-2$ D-flatness conditons (\ref{eq:FDN}) and (\ref{eq:DDN}).

Once the Fayet-Iliopoulos parameters $(\zeta_1,\cdots,\zeta_{N-1})$ are fixed,
relative size of $|p_{(i)}|^2$ are determined by (\ref{eq:DDN}),
and one obtains the following inequalities\footnote{We
ignore codimension one hyperplanes in the parameter space.},
\begin{equation}
|p_{(L_1)}|^2 < |p_{(L_2)}|^2 < \cdots < |p_{(L_N)}|^2.
\label{eq:ordering}
\end{equation}
Here $(L_1,L_2,\cdots,L_N)$ is a certain permutation of $(1,2,\cdots,N)$.

Next we rewrite the $2^N-2$ D-flatness conditions
by taking the ordering (\ref{eq:ordering}) into account\footnote{The rearrangement
of the D-flatness conditions is similar to that
considered in \cite{BGLP}.}.
They are classified into two types.
The first type of equations are written as
\begin{equation}
|p_{(L_1,\cdots,L_k)}|^2-|p_{(L_1,\cdots,L_k,L_{k+1})}|^2
-|p_{(L_1,\cdots,L_k,L_{k+2})}|^2+|p_{(L_1,\cdots,L_k,L_{k+1},L_{k+2})}|^2=0
\label{eq:D-flat-k}
\end{equation}
where $k=0,1,\cdots,N-2$.
The second type of equations are written as
\begin{equation}
|p_{(L_1,\cdots,L_i)}|^2-|p_{(a_1,\cdots,a_i)}|^2=\zeta_{(L_1,\cdots,L_i)-(a_1,\cdots,a_i)}
\label{eq:D-flat-i}
\end{equation}
where $i=1,2,\cdots,N-1$, and
$(a_1,\cdots,a_N)$ is a permutation of $(1,\cdots,N)$
which gives $p_{(a_1,\cdots,a_i)}$ other than $p_{(L_1,\cdots,L_i)}$.
Note that the number of independent equations in the second type is
\begin{equation}
\sum_{i=1}^{N-1} \left( (^N_{\, k})-1 \right) = 2^N-N-1.
\end{equation}
The parameter in the right hand side of the equation (\ref{eq:D-flat-i}) is defined by
\begin{equation}
\zeta_{(L_1,\cdots,L_i)-(a_1,\cdots,a_i)}
=\sum_{j=1}^i \zeta_{(L_j)-(a_j)},
\end{equation}
where $\zeta_{(L_i)-(a_i)}$ is a linear combination of $\zeta_i$ defined by
\begin{equation}
\zeta_{(L_j)-(a_j)}
=\left\{
\begin{array}{ll}
\zeta_{L_j}+\zeta_{L_j+1}+\cdots +\zeta_{a_j-1},
& ({\rm if} \; L_j<a_j)\\
-(\zeta_{a_j}+\zeta_{a_j+1}+\cdots +\zeta_{L_j-1}),
& ({\rm if} \; a_j<L_j)\\
0. & ({\rm if} \; L_j=a_j)\\
\end{array}
\right.
\end{equation}
By the definition of the permutation $(L_1,\cdots,L_N)$,
the parameter $\zeta_{(L_1,\cdots,L_i)-(a_1,\cdots,a_j)}$ turns out to be negative.
This implies that $|p_{(a_1,\cdots,a_i)}|^2$ is not zero,
and the variable $p_{(a_1,\cdots,a_i)}$ can be eliminated
by (\ref{eq:D-flat-i}) and the corresponding $U(1)$ gauge symmetry.
After the elimination of the redundant variables,
we obtain $N-1$ D-flatness conditions on
$N+1$ variables $p_{(L_1,\cdots,L_i)}$ $(i=0,\cdots,N)$,
\begin{equation}
|p_{(L_1,\cdots,L_k)}|^2-2|p_{(L_1,\cdots,L_k,L_{k+1})}|^2
+|p_{(L_1,\cdots,L_k,L_{k+1},L_{k+2})}|^2=-\zeta_{(L_{k+1})-(L_{k+2})},
\label{eq:DN}
\end{equation}
where $k=0,1,\cdots,N-2$.
By the definition of $(L_1,\cdots,L_N)$, the right hand side of 
(\ref{eq:DN}) is positive.
This implies that $\cal M$ is the resolution of the orbifold ${\bf C}^2/{\bf Z}_N$.
Thus the singularity of the orbifold is resolved in every region in the $(\zeta_1,\cdots,\zeta_{N-1})$
space (except for codimension one loci)
and non-geometric phases are projected out.

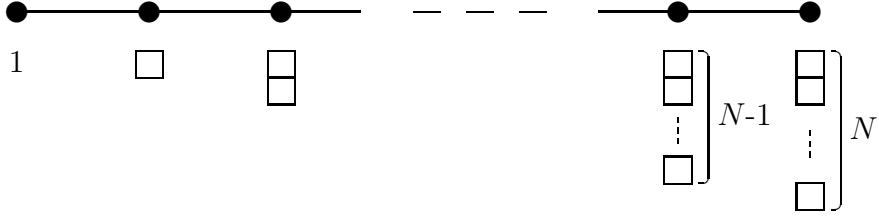
\begin{figure}[htp]
\begin{center}
\begin{picture}(400,120)
\put(47,77){1}
\put(95,75){\framebox(10,10)}
\put(145,75){\framebox(10,10)}
\put(145,65){\framebox(10,10)}
\put(295,75){\framebox(10,10)}
\put(295,65){\framebox(10,10)}
\put(295,35){\framebox(10,10)}
\put(308,60){\oval(7,50)[r]}
\put(315,57){$N$-1}
\put(345,75){\framebox(10,10)}
\put(345,65){\framebox(10,10)}
\put(345,25){\framebox(10,10)}
\put(358,55){\oval(7,60)[r]}
\put(365,52){$N$}
\put(50,100){\circle*{8}}
\put(100,100){\circle*{8}}
\put(150,100){\circle*{8}}
\put(300,100){\circle*{8}}
\put(350,100){\circle*{8}}
\put(200,100){\line(1,0){10}}
\put(220,100){\line(1,0){10}}
\put(240,100){\line(1,0){10}}
\put(300,50){\line(0,1){2}}
\put(300,54){\line(0,1){2}}
\put(300,58){\line(0,1){2}}
\put(350,45){\line(0,1){2}}
\put(350,49){\line(0,1){2}}
\put(350,53){\line(0,1){2}}
\thicklines
\put(50,100){\line(1,0){130}}
\put(270,100){\line(1,0){80}}
\end{picture}
\caption{Correspondence between representations of $SU(N)$
and the toric diagram for ${\bf C}^2/{\bf Z}_N$.}
\label{fig:representationC2ZN}
\end{center}
\end{figure}

Finally we would like to discuss correspondence between geometry of ${\bf C}^2/{\bf Z}_N$
and representation theory.
Similar argument to the $N=3$ case
shows that the $(^N_{\, k})$ vectors in $F_k$ corresponds to the
$k$-th vertex from the left in the toric diagram
and they form an antisymmetric representation of $SU(N)$ Lie algebra
as shown in Figure \ref{fig:representationC2ZN}.
By defining positivity of roots of $SU(N)$ Lie algebra in a similar manner to the last section,
we find out that the variables $p_{(L_1,\cdots,L_i)}$ remaining after the elimination of the redundant variables
correspond to the lowest weight state of each antisymmetric representation of $SU(N)$.

\section{Three-dimensional orbifolds}
\begin{figure}[htp]
\begin{center}
\begin{picture}(400,120)
\put(260,110){$n_3$}
\put(310,10){$n_1$}
\put(80,60){$n_2$}
\put(80,32){$\{n_4,\cdots,n_{10}\}$}
\put(140,18){$\{n_{11},\cdots,n_{24}\}$}
\put(200,4){$\{n_{25},\cdots,n_{31}\}$}
\put(125,38){\vector(1,1){18}}
\put(180,24){\vector(1,2){16}}
\put(235,10){\vector(1,3){14}}
\put(100,60){\circle*{8}}
\put(150,60){\circle*{8}}
\put(200,60){\circle*{8}}
\put(250,60){\circle*{8}}
\put(250,110){\circle*{8}}
\put(300,10){\circle*{8}}
\thicklines
\put(100,60){\line(1,0){150}}
\put(100,60){\line(3,1){150}}
\put(100,60){\line(4,-1){200}}
\put(150,60){\line(2,1){100}}
\put(150,60){\line(3,-1){150}}
\put(200,60){\line(1,1){50}}
\put(200,60){\line(2,-1){100}}
\put(250,60){\line(0,1){50}}
\put(250,60){\line(1,-1){50}}
\put(250,110){\line(1,-2){50}}
\end{picture}
\caption{Toric diagram for ${\bf C}^3/{\bf Z}_7$.}
\label{fig:toricC3Z7}
\end{center}
\end{figure}
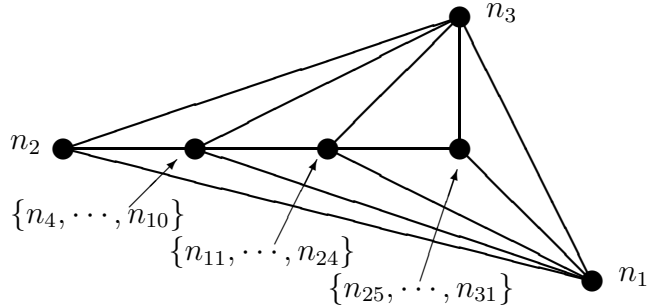

\reseteqnum
In this section we consider D-branes  on the orbifold ${\bf C}^3/{\bf Z}_N$
where the action of $g \in {\bf Z}_N$ on $(x,y,z) \in {\bf C}^3$ is given by
\begin{equation}
(x, y, z) \rightarrow (\omega_N x, \omega_N y,\omega_N^{-2}z),
\quad \omega_N={\rm exp}(2\pi i/N).
\label{eq:ZN}
\end{equation}
By a similar analysis to the two-dimensional case,
we obtain toric data of the vacuum moduli space of the quiver gauge theory.
As an example, we consider the orbifold ${\bf C}^3/{\bf Z}_7$.
F-flatness conditions are represented by a cone $\hat \sigma$ in ${\bf R}^9$ generated
by 21 vectors $m_i$,
and its dual cone $\sigma$ turns out to be generated by 31 vectors $n_i$.
(The form of the vectors is given in the Appendix.)
After combining D-flatness conditions,
we obtain the toric diagram depicted in Figure \ref{fig:toricC3Z7}.
In the toric diagram
7 vectors $n_4,\cdots,n_{10}$ correspond to the left vertex inside the diagram,
14 vectors $n_{11},\cdots,n_{24}$ correspond to the middle one
and 7 vectors $n_{25},\cdots,n_{31}$ correspond to the right one.

The structure of the multiplicity can be analyzed from the explicit form of the
nine-dimensional vectors $n_i$.
However, not all the components of the vector are necessary
to describe the structure of the multiplicity as in the two-dimensional case.
In this case it is enough to consider only seven components of $n_i$, which we will denote by $\hat n_i$.
The 7 vectors corresponding to the left vertex inside the toric diagram are given by the columns of the
matrix,
\begin{equation}
\left(
\begin{array}{ccccccc}
\hat n_4&\hat n_5&\hat n_6&\hat n_7&\hat n_8&\hat n_9&\hat n_{10}\\
1&0&0&0&0&0&0\\
0&1&0&0&0&0&0\\
0&0&1&0&0&0&0\\
0&0&0&1&0&0&0\\
0&0&0&0&1&0&0\\
0&0&0&0&0&1&0\\
0&0&0&0&0&0&1
\end{array}
\right),
\label{eq:left}
\end{equation}
the 14 vectors corresponding to the middle vertex inside the toric diagram
are given by the columns of the matrix,
\begin{equation}
\left(
\begin{array}{cccccccccccccc}
\hat n_{11}&\hat n_{12}&\hat n_{13}&\hat n_{14}&\hat n_{15}&\hat n_{16}&\hat n_{17}&
\hat n_{18}&\hat n_{19}&\hat n_{20}&\hat n_{21}&\hat n_{22}&\hat n_{23}&\hat n_{24}\\
1&0&0&0&0&1&0&1&0&0&0&1&0&0\\
0&1&0&0&0&0&1&0&1&0&0&0&1&0\\
1&0&1&0&0&0&0&0&0&1&0&0&0&1\\
0&1&0&1&0&0&0&1&0&0&1&0&0&0\\
0&0&1&0&1&0&0&0&1&0&0&1&0&0\\
0&0&0&1&0&1&0&0&0&1&0&0&1&0\\
0&0&0&0&1&0&1&0&0&0&1&0&0&1
\end{array}
\right),
\label{eq:middle}
\end{equation}
and the 7 vectors corresponding to the right vertex inside the toric diagram
are given by the columns of the matrix,
\begin{equation}
\left(
\begin{array}{ccccccc}
\hat n_{25}&\hat n_{26}&\hat n_{27}&\hat n_{28}&\hat n_{29}&\hat n_{30}&\hat n_{31}\\
1&0&0&1&0&1&0\\
0&1&0&0&1&0&1\\
1&0&1&0&0&1&0\\
0&1&0&1&0&0&1\\
1&0&1&0&1&0&0\\
0&1&0&1&0&1&0\\
0&0&1&0&1&0&1
\end{array}
\right).
\label{eq:right}
\end{equation}

\begin{figure}[htdp]
\begin{center}
\leavevmode
\epsfysize=50mm
\epsfbox{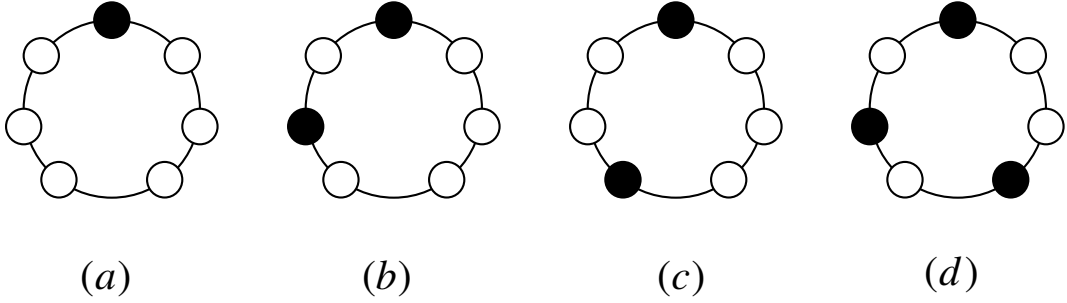}
\caption{Diagrams representing the structure of $\hat n_i$ for ${\bf C}^3/{\bf Z}_7$.
The black circles represent positions where the entries are equal to 1.
$(a)$ represents the vectors $\{ \hat n_4,\cdots,\hat n_{10} \}$ corresponding to the
left vertex inside the toric diagram.
$(b)$ and $(c)$ represent the vectors $\{ \hat n_{11},\cdots,\hat n_{24} \}$
corresponding to the middle vertex inside the toric diagram.
$(d)$ represents vectors $\{ \hat n_{25},\cdots,\hat n_{31} \}$ corresponding to
the right vertex inside the toric diagram.}
\label{fig:graph}
\end{center}
\end{figure}

The structure of these vectors can be represented diagrammatically
as in Figure \ref{fig:graph}.
The seven small circles in the graph correspond to the seven components of each
vector $\hat n_i$
and the black circles represent the positions where the entries are equal to 1.
Each diagram represents seven vectors by performing ${\bf Z}_7$ rotations.
Thus the vectors on the $k$-th vertex from the left inside the toric diagram
are obtained by  choosing $k$ positions from seven
with the constraints that two neighboring positions must not be chosen.

This structure also holds for the orbifold ${\bf C}^3/{\bf Z}_N$
defined by (\ref{eq:ZN}),
and the total number $n_D(N)$ of the vectors $n_i$ necessary to describe the D-geometry
of the orbifold is given by
\begin{equation}
n_D(N)=\sum_{k=1}^{[N/2]} \frac{N}{N-k}
\left(
\begin{array}{c}
N-k \\
k
\end{array}
\right)+3.
\end{equation}
Note that this equation also holds when $N$ is even.
If we define
\begin{equation}
f(N)=n_D(N)-2,
\end{equation}
$f(N)$ satisfies the recursion relation similar to the Fibonacci sequence,
\begin{equation}
f(N)=f(N-1)+f(N-2)
\end{equation}
with the conditions $f(1)=1$ and $f(2)=3$.
The solution of the recursion relation is given by
\begin{equation}
f(N)=\left(\frac{1+\sqrt 5}{2}\right)^N +\left(\frac{1-\sqrt 5}{2}\right)^N,
\end{equation}
and hence analytic expression of $n_D(N)$ is
\begin{equation}
n_D(N)=\left(\frac{1+\sqrt 5}{2}\right)^N +\left(\frac{1-\sqrt 5}{2}\right)^N+2.
\end{equation}
It reproduces the numbers given in the Table in the Introduction.

\vskip 1cm
\centerline{\large\bf Acknowledgements}

I would like to thank T. Tani for valuable discussions.

\appendix

\section{Toric data for the orbifold ${\bf C}^3/{\bf Z}_7$.}
\reseteqnum

\begin{equation}
K=\left(
\begin{array}{cccccccccc}
m_1&1&0&0&0&0&0&0&0&0\\
m_2&0&1&0&0&0&0&0&0&0\\
m_3&0&0&1&0&0&0&0&0&0\\
m_4&0&0&0&1&0&0&0&0&0\\
m_5&0&0&0&0&1&0&0&0&0\\
m_6&0&0&0&0&0&1&0&0&0\\
m_7&0&0&0&0&0&0&1&0&0\\
m_8&0&0&0&0&0&0&0&1&0\\
m_9&0&0&0&0&0&0&0&0&1\\
m_{10}&1&0&1&0&-1&0&0&0&0\\
m_{11}&1&0&1&1&-1&-1&0&0&0\\
m_{12}&1&0&1&1&0&-1&-1&0&0\\
m_{13}&1&0&1&1&0&0&-1&-1&0\\
m_{14}&1&0&1&1&0&0&0&-1&-1\\
m_{15}&1&0&0&1&0&0&0&0&-1\\
m_{16}&0&1&1&0&0&0&0&0&-1\\
m_{17}&0&1&0&1&0&0&0&0&-1\\
m_{18}&0&1&0&0&1&0&0&0&-1\\
m_{19}&0&1&0&0&0&1&0&0&-1\\
m_{20}&0&1&0&0&0&0&1&0&-1\\
m_{21}&0&1&0&0&0&0&0&1&-1
\end{array}
\right)
\end{equation}

\[
T=\left(
\begin{array}{ccccccccccccccccc}
n_1&n_2&n_3&n_4&n_5&n_6&n_7&n_8&n_9&n_{10}&
n_{11}&n_{12}&n_{13}&n_{14}&n_{15}&n_{16}&n_{17}\\
0&1&0&0&0&1&1&1&1&1&0&0&1&1&1&0&0\\
0&0&1&0&0&0&0&0&0&1&0&0&0&0&1&0&1\\
1&0&0&1&0&0&0&0&0&0&1&0&0&0&0&1&0\\
1&0&0&0&1&0&0&0&0&0&0&1&0&0&0&0&1\\
1&0&0&0&0&1&0&0&0&0&1&0&1&0&0&0&0\\
1&0&0&0&0&0&1&0&0&0&0&1&0&1&0&0&0\\
1&0&0&0&0&0&0&1&0&0&0&0&1&0&1&0&0\\
1&0&0&0&0&0&0&0&1&0&0&0&0&1&0&1&0\\
1&0&0&0&0&0&0&0&0&1&0&0&0&0&1&0&1
\end{array}
\right.
\nonumber
\]

\begin{equation}
\left.
\begin{array}{ccccccccccccccc}
n_{18}&n_{19}&n_{20}&n_{21}&n_{22}&n_{23}&n_{24}&
n_{25}&n_{26}&n_{27}&n_{28}&n_{29}&n_{30}&n_{31}\\
0&0&1&1&0&0&1&0&0&1&0&0&0&0\\
0&0&0&1&0&0&1&0&0&1&0&1&0&1\\
1&0&0&0&1&0&0&1&0&0&1&0&1&0\\
0&1&0&0&0&1&0&0&1&0&0&1&0&1\\
0&0&1&0&0&0&1&1&0&1&0&0&1&0\\
1&0&0&1&0&0&0&0&1&0&1&0&0&1\\
0&1&0&0&1&0&0&1&0&1&0&1&0&0\\
0&0&1&0&0&1&0&0&1&0&1&0&1&0\\
0&0&0&1&0&0&1&0&0&1&0&1&0&1
\end{array}
\right)
\end{equation}

\end{document}